\documentclass[pra,twocolumn,showpacs,amsmath,amssymb,superscriptaddress,longbibliography]{revtex4-1}
\usepackage{psfrag,graphicx}
\usepackage{amsfonts,amssymb,amsmath}      
\usepackage{graphicx}
\usepackage{dcolumn}
\usepackage{bm}
\usepackage{float}

\usepackage{graphicx}

\usepackage{comment}

\usepackage{hyperref}
\usepackage{accents}
\usepackage{bbding}
\usepackage{color}
\usepackage{soul}
\usepackage{physics}
\usepackage{epstopdf}
\usepackage{changepage}
\usepackage{orcidlink}
\usepackage[normalem]{ulem}
\usepackage{dsfont}
\usepackage[scr=boondoxo]{mathalpha}

\usepackage{xcolor}

\renewcommand{\erf}[1]{Eq.~(\ref{#1})}
\newcommand{\beq}{\begin{equation}}
\newcommand{\eeq}{\end{equation}}

\newcommand{\dg}{^\dagger}

\newcommand{\smallfrac}[2]{\mbox{$\frac{#1}{#2}$}}

\newcommand{\half}{\smallfrac{1}{2}}
\renewcommand{\op}[2]{\ketbra{#1}{#2}}
\newcommand{\sch}{Schr\"odinger}

\newcommand{\sq}[1]{\left[ {#1} \right]}

\renewcommand{\Tr}{\text{Tr}}

\newcommand{\ex}[1]{\langle{#1}\rangle}
\renewcommand{\dd}{{\rm d}}

\newcommand{\ie}{{\em i.e.}}

\newcommand{\inv}{^{-1}}

\renewcommand{\sq}{_{\cal S}}
\renewcommand{\ng}{_{\cal NG}}
\newcommand{\ngmx}{\tilde{E}_{\mathcal{N}\mathcal{G}}}
\newcommand{\ha}{\hat{a}}

\newcommand{\incoh}{_{\cal I}}
\newcommand{\mx}{_{{\cal T}h}}
\newcommand{\coh}{_{\cal C}}

\renewcommand{\bar}[1]{\overline{#1}}

\definecolor{nblue}{rgb}{0.06,0.3,0.73}
\definecolor{nblack}{rgb}{0,0,0}
\definecolor{nred}{rgb}{0.9,0.1,0.1}
\definecolor{nmagenta}{rgb}{0.7,0.0,0.3}
\definecolor{applegreen}{rgb}{0.55, 0.71, 0.0}
\definecolor{nteal}{rgb}{0, 0.8, 0.8}

\newcommand\stPW{\bgroup\markoverwith{\applegreen{\rule[0.5ex]{2pt}{0.4pt}}}\ULon}

\begin{document}


\title{Energetics of non-Gaussianity in single mode cavities}
\author{Sahil S. Jafar\orcidlink{0009-0007-2011-3669}}
\email{sahilsardarjafar@gmail.com}
\affiliation{MajuLab, CNRS-UCA-SU-NUS-NTU International Joint Research Laboratory}
\affiliation{Centre for Quantum Technologies, National University of Singapore, 117543 Singapore, Singapore}
\affiliation{School of Physical and Mathematical Sciences, Nanyang Technological University, 21 Nanyang Link, Singapore 637371, Singapore}
\author{Kian Hwee Lim\orcidlink{0000-0003-2154-4288}}
\email{kianhwee.lim95\@gmail.com}
\affiliation{MajuLab, CNRS-UCA-SU-NUS-NTU International Joint Research Laboratory}
\affiliation{Centre for Quantum Technologies, National University of Singapore, 117543 Singapore, Singapore}
\affiliation{CNRS@CREATE LTD, 1 Create Way, \#08-01 CREATE Tower, Singapore 138602}
\author{Kiarn T. Laverick\orcidlink{0000-0002-3688-1159}}
\email{kiarn.l@nus.edu.sg}
\affiliation{MajuLab, CNRS-UCA-SU-NUS-NTU International Joint Research Laboratory}
\affiliation{Centre for Quantum Technologies, National University of Singapore, 117543 Singapore, Singapore}
\author{Samyak P. Prasad\orcidlink{0009-0001-3672-6001}}
\affiliation{MajuLab, CNRS-UCA-SU-NUS-NTU International Joint Research Laboratory}
\affiliation{Centre for Quantum Technologies, National University of Singapore, 117543 Singapore, Singapore}
\author{Maria Maffei\orcidlink{0000-0001-5183-4716}}
\affiliation{Universit\'e de Lorraine, CNRS, LPCT, F-54000 Nancy, France}

\author{Alexia Auff\`eves\orcidlink{0000-0003-4682-5684}}
\email{alexia.auffeves@cnrs.fr}
\affiliation{MajuLab, CNRS-UCA-SU-NUS-NTU International Joint Research Laboratory}
\affiliation{Centre for Quantum Technologies, National University of Singapore, 117543 Singapore, Singapore}

\date{\today}

\begin{abstract}
Non-Gaussian states are key resources for quantum technologies, making the quantification of non-Gaussianity a fundamental challenge. We introduce an energetic framework for characterizing non-Gaussianity in single-mode bosonic states by decomposing the total energy into Gaussian and non-Gaussian contributions. For pure states, we show that the non-Gaussian energy defines a bona fide measure of non-Gaussianity and establish a direct connection with the relative entropy of non-Gaussianity. As an illustration, we consider non-Gaussian states generated from coherent states with tunable amplitudes using a SNAP gate. We find that the resulting non-Gaussian energy and Wigner negativity are maximized at similar input amplitudes.
 For mixed states, we demonstrate that the non-Gaussian energy provides a faithful witness of non-Gaussianity. Our results uncover an energetic fine structure of non-Gaussian quantum states and offer new insights into the efficient generation and manipulation of non-Gaussian resources.

\end{abstract}
\pacs{}
\maketitle

\section{Introduction}

Bosonic systems, such as single mode cavities, are simple textbook systems of quantum physics. As they support states with a clear classical limit, they provide appealing platforms to explore the quantum-classical boundary through, for example, the generation of Schr\"odinger cat states and the study of their decoherence \cite{decoherence_serge,haroche2006exploring,he2023fast}, measurement theory~\cite{WisemanMilburn2010}, and the emergence of classicality~\cite{Quantum-To-ClassicalTransitionSchlosshauer2007,Zurek}. They have also emerged as a promising paradigm for quantum technologies, especially in quantum computing with continuous variables \cite{Lloyd_Computation,chabaud2023resources,upreti2026bounding,CV_qi_Braunstein}, quantum communication \cite{usenko2026continuous,joshi2021quantum,weedbrook}, and quantum sensing~\cite{quantum_sensing_1}.

Non-Gaussian states are key to capture the features of non-classical physics, and have been shown to lead to quantum advantages~\cite{Eisert_distillation,no-go-theorem-gaussian-error,ahmed2026nogotheoremgaussianquantum,Mari_eisert,Veitch_2012_negative_prob,nonlocality_wigner, Contextuality_and_Wigner}, making their detection and quantification a crucial ability for both foundations and applications. Accordingly, several measures of non-Gaussianity have been proposed. These include information-theoretic measures such as relative entropy of non-Gaussianity~\cite{non-gauss_2,nongauss_1,relative_entropy_marian_exact,Faithful_measure_from_reng} and Hilbert–Schmidt distance~\cite{HIlbeert_schmidt_meausre}, phase-space based measures such as Wigner negativity~\cite{negativity_measure, non_gauss_0} and maximum negentropy of quadrature distributions~\cite{negentropy_based}, as well as operational measures such as the robustness of non-Gaussianity~\cite{robustness-1,robustness-2}. Experimentally motivated measures include stellar rank~\cite{stellar_rank} and non-Gaussian control parameters~\cite{fumiya_Hanamura_paper_ng}.

Recently, energy has emerged as an interesting physical quantity to measure and detect quantum features such as entanglement, both in continuous variables~\cite{gerardo_ergotrpic_cv, Friis_energetics_correlations, Serafini_entanglement_symplectic, eisert_optimal_entnaglement_witness_2006} and in discrete variables~\cite{laverick2026energetic,Alimuddin_Yang2024_ergotrpic_entangleemnt,Alimuddin-2,Perarnau-Llobet_correlations_work}. Energetic measures of quantum resources are constrained by conservation laws, providing a direct route to understanding fundamental trade-offs~\cite{laverick2026energetic}. Moreover, energetic quantities offer operational advantages, since they can be determined without necessarily requiring the knowledge of the full quantum state. They have recently been proposed as cost-functions to optimize the generation of non-Gaussian states in the scattering of coherent light by a two-level atom~\cite{lim2026energetic}.

In this work, we decompose the total energy of a pure single-mode bosonic state into energetic contributions associated with displacement, squeezing, and non-Gaussianity.  We show that the non-Gaussian contribution to energy, which we dub the \emph{non-Gaussian energy}, is a measure of non-Gaussianity which can be easily accessed in experiments. We further establish its connection with a well known entropic measure, namely, the relative entropy of non-Gaussianity. 
As an example, we study the non-Gaussian energy of states generated by SNAP operations~\cite{snap1} on coherent states and compare it with Wigner negativity~\cite{negativity_measure}, finding that both are maximized at nearly the same initial-state parameter.
 Switching to mixed states, we introduce a similar energetic decomposition, but modified by an additional contribution arising from the mixedness of the state. Quantifying the mixedness by the energy of a thermal state with the same entropy, we find that the modified non-Gaussian contribution serves as a faithful energetic witness of non-Gaussianity.

The work is organized as follows. Section~\ref{Sec:NG_Energy_Pure} introduces the energetic decomposition and non-Gaussian energy for pure states. Section~\ref{sec3: Optimising non-Gaussianity under energy constraints} illustrates the usefulness of the framework and examines non-Gaussianity generated by SNAP operations. Finally, Sec.~\ref{sec:Extension to mixed states} extends the framework to mixed states.

\section{Non-Gaussian energy for pure states}\label{Sec:NG_Energy_Pure}
\subsection{Energetic decomposition for pure states}
\begin{figure}[h]
     \centering
     \includegraphics[width=1\linewidth]{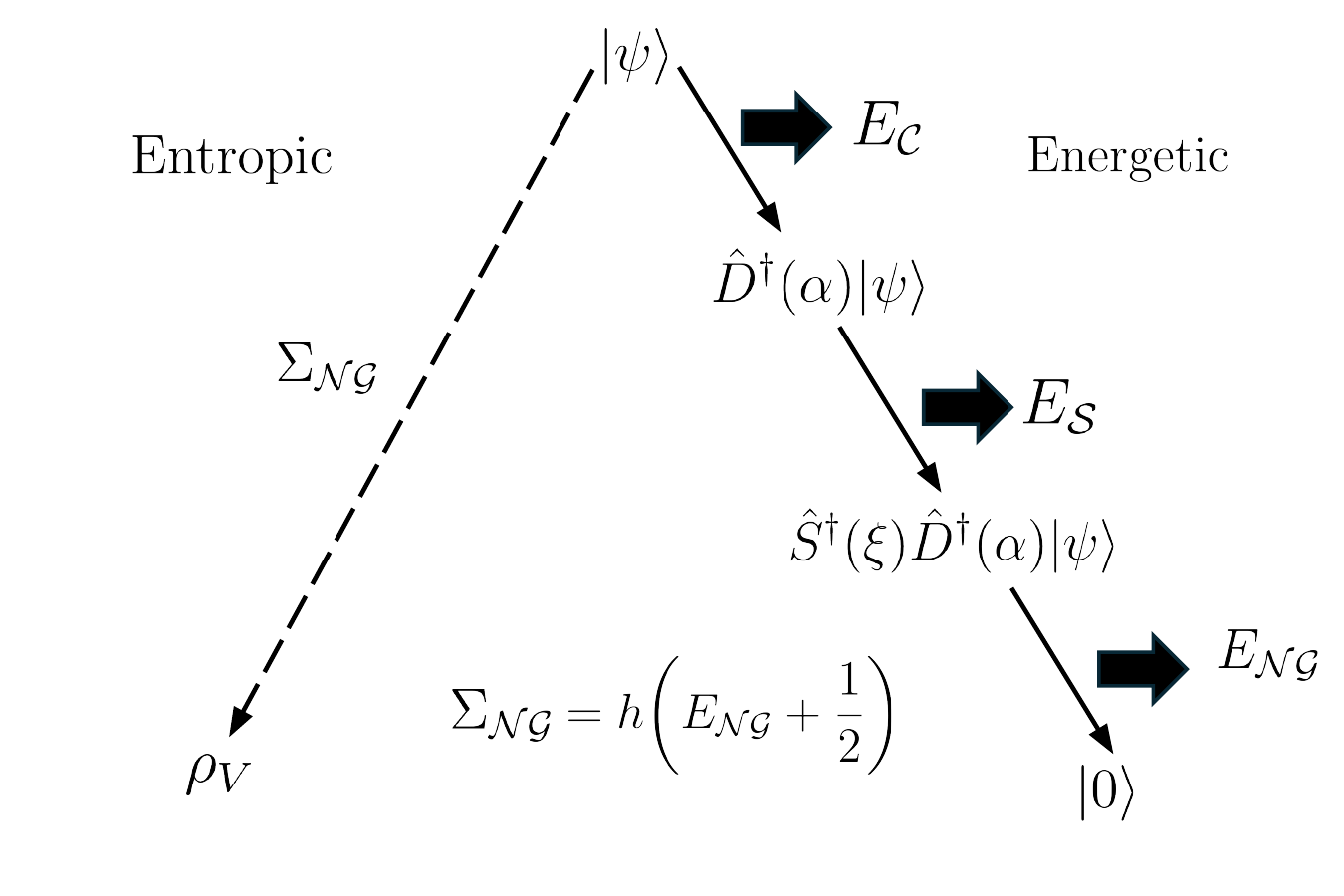}
     \caption{A diagram of our energetic approach to quantifying non-Gaussianity compared to the standard entropic approach. In our energetic approach, we systematically extract energy through Gaussian unitaries (displacements and squeezing) and quantify the remaining energy of the state extractable only through non-Gaussian unitaries as our measure of non-Gaussianity.  The states we obtain thereby have the same entropy but different energies. In contrast, the entropic approach involves relative entropy to the reference Gaussian state $\rho_V$, which has same energy but different entropy. We have shown that these two approaches are equivalent for pure states, since the entropic measure is related to $E\ng$ through the function $h$ [see \erf{h(x)}].}
     \label{fig:enrgetic-diagram-pure}
 \end{figure}

We consider a single bosonic mode described by the annihilation and creation operators $\ha$ and $\ha^\dagger$, satisfying the commutation relation $[\ha,\ha^\dagger]=1$. The total energy of the system in state $\ket{\psi}$ is counted in number of photons and reads $E = \ex{ \ha^\dagger \ha }$. The primary question we consider here is the following: can we distinguish, purely from an energetic perspective, whether a  pure state $\ket{\psi}$ is Gaussian, that is, whether its Wigner function is a Gaussian distribution~\cite{ Serafini2017,weedbrook}? Gaussian states are completely specified by their  first moment $\ex{\ha}$, and their second moments, which are collected in the covariance matrix $V$, given by
\begin{equation}
V =
\left(\langle \delta \ha^\dagger \delta \ha \rangle+\frac{1}{2}  \right)I_2
+
\begin{pmatrix}
\mathrm{Re}\,\langle (\delta \ha)^2 \rangle\, & \mathrm{Im}\,\langle (\delta \ha)^2 \rangle\, \\
\mathrm{Im}\,\langle (\delta \ha)^2 \rangle\, & -\mathrm{Re}\,\langle (\delta \ha)^2 \rangle
\end{pmatrix}\,, \label{covariance matrix}
\end{equation}
where $\delta \hat{O} = \hat{O} - \ex{\hat{O}}$ and $I_2$ denotes the $2\times2$ identity matrix. To answer the question, we decompose the total energy as
\begin{equation}
    E = E\coh + E\sq+E\ng\,, \label{eq:energetic_decomposition_pure}
\end{equation}
where 
we define
\begin{align}
    E\coh &:= |\ex{\hat{a}}|^2 \label{eq:coherent energy}\,,\\
    E\sq &:= \ex{\delta \ha ^\dagger \delta \ha} - \bigg(\sqrt{\det V}-\frac{1}{2}\bigg)\,, \label{eq:squeezing energy}\\
    E\ng &:= {\sqrt{\det V}}-\frac{1}{2}\,.\label{eq:non-Gaussian energy_pure}
\end{align}
The \emph{coherent energy} $E\coh$, captures the maximum amount of energy that can be extracted from the state $\ket{\psi}$ using displacement unitaries. Specifically, one can extract $E\coh$ by applying the displacement operator
$\hat{D}^\dagger(\alpha)=e^{\alpha^* \ha-\alpha \ha^\dagger}$, 
with amplitude
$\alpha=\ex{\ha}.$ The extraction of coherent energy is depicted in Fig.~\ref{fig:enrgetic-diagram-pure}. In the standard energetic decomposition based on the mean-field decomposition~\cite{laverick2026energetic, prasad2026thermodynamics}, the energy remaining in the state after deducting the coherent energy is known as the \emph{incoherent energy}, $E\incoh = E-E\coh =\ex{\delta \ha ^\dagger \delta \ha }$, which is the energy associated with the quantum fluctuations about the
mean field. Here, we have further resolved the incoherent energy  into two parts, \ie, $E\incoh = E\sq+E\ng$. 

$E\sq$  corresponds to the maximum amount of energy extractable from the state $\hat{D}^\dagger(\alpha)\ket{\psi}$ using squeezing unitaries. It has been proved in~\cite{marcus_passive,vittorio_work_extraction} that the optimal squeezing unitary is $\hat{S}^\dagger(\xi)=e^{\xi\ha^2-\xi^* {\ha^\dagger}^2}$, where the squeezing parameter $\xi$ is chosen such that the covariance matrix is transformed to its Williamson form, \ie,
$\sqrt{\det V}\,{I}_2$~\cite{Williamson_theorem_paper_1936}. 
It is particularly illustrative to express $E\sq$ in terms of the principal quadratures,
$\hat{q}_\phi = \ha^\dagger e^{i\phi}+\ha e^{-i\phi}$ and
$\hat{p}_\phi = i(\ha^\dagger e^{i\phi}- \ha e^{-i\phi})$,
where the phase $\phi$ is chosen so that the covariance matrix is diagonal. In this representation, $E\sq$ takes the following form:
\begin{equation}
    E\sq = \frac{1}{4}\bigg( \sqrt{\ex{(\delta\hat{q}_{\phi})^2}}  -\sqrt{\ex{(\delta\hat{p}_{\phi})^2}}\bigg)^2\,. \label{squeezing energy_rotated}
\end{equation}
The above equation shows that $E\sq$ is a non-negative quantity and originates from the difference between the standard deviations of the principal quadratures. Importantly, the squeezed energy is zero for isotropic states, \ie, states satisfying $\ex{\delta\hat{q}_{\phi}^2}=\ex{\delta\hat{p}_{\phi}^2}$. Following the naming convention established by the coherent energy to name the energetic contribution after the state that only contains this energy (\ie, coherent states), we refer to $E\sq$ as the {\em squeezed energy}, since squeezed vacuum states contain only this energetic contribution. This further extraction of the squeezed energy is depicted in Fig.~\ref{fig:enrgetic-diagram-pure}.

We have accounted for the energetic contribution arising due to displacement and squeezing which are single mode Gaussian unitaries. The resulting state, $\hat{S}^\dagger(\xi)\hat{D}^\dagger(\alpha)\ket{\psi}$ is the minimum-energy state that can be prepared from $\ket{\psi}$ using Gaussian unitaries, and is called the Gaussian-passive state~\cite{marcus_passive,vittorio_work_extraction}. The order of displacement and squeezing is important for our energetic decomposition. If squeezing is applied before displacement, the extracted energies do not correspond to $E\coh$ and $E\sq$ individually. Nevertheless, the total, \ie,  $E\coh + E\sq$, is the maximum energy that can be extracted via Gaussian unitaries. Note that the Gaussian-passive state is not necessarily a core state which is defined in context of the stellar function as elaborated in Appendix~\ref{appendix:core state}. 

For a pure Gaussian state, its corresponding Gaussian-passive state is the vacuum, meaning that the coherent and squeezed energies account for all the energy in the state. In contrast, for a non-Gaussian state, the corresponding Gaussian-passive state is not the vacuum state and has an energy equal to $\sqrt{\det V}-\tfrac{1}{2}$, which we identify as the \emph{non-Gaussian energy}, $E\ng$.  Finally, the removal of this energy from the Gaussian-passive state, which necessarily requires a non-Gaussian unitary, eliminates all remaining energy, thereby leaving the state in the vacuum (see Fig.~\ref{fig:enrgetic-diagram-pure}).

\subsection{Non-Gaussian energy as a measure of non-Gaussianity}
The central result of our work is that  $E\ng$ is a measure of non-Gaussianity for pure states. To show this, we must demonstrate the following properties of $E\ng$~\cite{non_gauss_0}:
\begin{enumerate}
\item {\it Positivity}---It is a well-defined, non-negative quantity, \ie, $E\ng \geq 0$. \label{prop:1_pure}
    \item {\it Faithfulness}---$E\ng = 0$, iff $\ket{\psi}$ is a pure Gaussian state.\label{prop:2_pure}
    \item {\it Invariance under Gaussian unitaries}---$E\ng[\ket{\psi}] = E\ng[\hat{U}_G\ket{\psi}]$ for any Gaussian unitary $\hat{U}_G$, \ie, any unitary of the form $\hat{U}_G=e^{i\hat{G}}$ where $\hat{G}$ is a Hermitian operator at most quadratic in the annihilation and creation operators.\label{prop:3_pure}
\end{enumerate}
Properties~\ref{prop:1_pure} and~\ref{prop:2_pure} follow directly from the Robertson-\sch{} uncertainty relation~\cite{weedbrook,Serafini2017}, which states that,
\begin{equation}
    {\det V} \geq \frac{1}{4}\,. \label{eq:eng_pure_varxvarp}
\end{equation}
The equality is satisfied iff the state is a pure Gaussian, in which case, $E\ng=0$. To prove property~\ref{prop:3_pure}, we use the fact that Gaussian unitaries preserve the symplectic eigenvalues of the covariance matrix~\cite{Serafini2017,weedbrook}. For a single mode covariance matrix,  $\sqrt{\det V}$ is the only symplectic eigenvalue, which implies that $E\ng$ is invariant under Gaussian unitaries. Thus, properties~\ref{prop:1_pure}, \ref{prop:2_pure}, and \ref{prop:3_pure} make $E\ng$ a valid measure of non-Gaussianity for pure states~\cite{non_gauss_0}.

This result motivates the investigation of the relationship between $E\ng$ and another well-known measure of non-Gaussianity, namely the relative entropy of non-Gaussianity, $\Sigma\ng[\rho]$~\cite{non_gauss_0,nongauss_1,non-gauss_2,relative_entropy_marian_exact}. It is defined as
\begin{equation}
    \Sigma\ng[\rho]
    = \min_{\sigma \in {\mathbb G}}D[\rho ||\sigma]\,.
    \label{eq:rel_entropy_nongauss}
\end{equation}
${\mathbb G}$ denotes the set of Gaussian states and the quantum relative entropy is given by $D[\rho_1 || \rho_2]:= \Tr\!\left[\rho_1(\ln \rho_1 - \ln \rho_2)\right]$.  Computing the minimum over the set of Gaussian states in Eq.~\eqref{eq:rel_entropy_nongauss} yields the simpler expression~\cite{relative_entropy_marian_exact}
\begin{equation}
\Sigma\ng[\rho] = S[\rho_V] - S[\rho],
\label{relative enrtorpy-3}
\end{equation}
where $S[\rho] = -\mathrm{Tr}[\rho\ln\rho]$ is the von Neumann entropy of the state $\rho$. $\rho_V$, referred to as the \emph{reference Gaussian state}, is the Gaussian state with the same first moments and covariance matrix as $\rho$. Since the quantum relative entropy is non-negative, the reference Gaussian state has the maximum von Neumann entropy among all states having the same first moment and covariance matrix, reflecting the extremality property of Gaussian states~\cite{extremality_gaussian}.

Our energetic approach quantifies non-Gaussianity by comparing a state with the Gaussian state sharing its entropy, whereas the relative entropy of non-Gaussianity compares it with the Gaussian state sharing its energy, as illustrated in Fig.~\ref{fig:enrgetic-diagram-pure}. These two approaches therefore provide complementary energetic and entropic perspectives on non-Gaussianity. Going further, we show that they become fully equivalent for pure states. Indeed, the von Neumann entropy of the Gaussian state $\rho_V$ reads~\cite{Demarie_2018,Serafini2017} \begin{equation} S[\rho_V] = h(\sqrt{\det V}),, \label{eq:S of rhoV} \end{equation} 
where 
\begin{equation}     h(x) = \bigg(x+\frac{1}{2}\bigg)\ln{\bigg(x+\frac{1}{2}\bigg)} - \bigg(x-\frac{1}{2}\bigg)\ln{\bigg(x-\frac{1}{2}\bigg)}\,.\label{h(x)} \end{equation} 
For a pure state $\rho=\op{\psi}{\psi}$, for which $S[\rho]=0$, the relative entropy of non-Gaussianity reduces to 
\begin{equation}    \Sigma\ng[\op{\psi}{\psi}] = h(E\ng+1/2)\,.
\end{equation} 
To establish the relation between $\Sigma\ng$ and $E\ng$, we note that the derivative of $h(x)$ is 
\begin{equation} h'(x)=\ln\left(\frac{2x+1}{2x-1}\right). \end{equation} 
Since $h'(x)\geq 0$ for all $x\in[1/2,\infty)$, the function $h(x)$ is monotonically increasing. Consequently, $\Sigma\ng$ is a monotonic function of $E\ng$: increasing the non-Gaussian energy necessarily increases the relative entropy of non-Gaussianity. We therefore conclude that, for pure single-mode bosonic states, $E\ng$ and $\Sigma\ng$ are equivalent quantifiers of non-Gaussianity, establishing a direct bridge between the energetic and entropic viewpoints.

\section{Maximising non-Gaussianity in pure bosonic states}\label{sec3: Optimising non-Gaussianity under energy constraints}
For a fixed total energy $E$, the maximum possible value of non-Gaussian energy is $E\ng = E$. This maximum is attained for states with $\langle \ha \rangle = 0$ and $\langle \ha^2 \rangle = 0$. 
To show this, we express $\det V$ using Eq.~\eqref{covariance matrix} as 
\begin{equation}
    \det V = \bigg(\langle\delta \ha^\dagger \delta \ha\rangle+\frac{1}{2}\bigg)^2 - |\langle (\delta \ha)^2\rangle|^2\,,
\end{equation}
from which we obtain the inequality
\begin{equation}
    \det V \leq \bigg(\langle\delta \ha^\dagger \delta \ha\rangle+\frac{1}{2}\bigg)^2  \leq \bigg(\langle\ha^\dagger \ha\rangle+\frac{1}{2}\bigg)^2\,, \label{inequality_1}
\end{equation}
Here, the first inequality is saturated when $\langle (\delta \ha)^2\rangle =0$ and the second inequality is saturated when $\ex{\ha}=0$. Rearranging Eq.~\eqref{inequality_1}, we obtain
\begin{equation}
    E\ng \leq  E\incoh \leq E\,.\label{eng_ei_e_inequality}
\end{equation}
Thus, $E\ng = E$ is satisfied when both the first and second inequalities are saturated, \ie, when $\langle \ha^2\rangle =0$ and $\ex{\ha}=0$.
We can identify a class of pure states with exactly this property: a superposition of Fock states $\ket{n}$ which differ from each other by at least three excitations, \ie, $\ket{\psi} =  \sum_{n \in \mathbb{K}} c_n \ket{n}$, where $\mathbb{K}$ is the set of non-negative integers such that $|n'-n''|\geq 3 \quad \forall \quad n' \neq n''  \in \mathbb{K}$. These states are ``maximally non-Gaussian" for a fixed energy according to our measure. Since the non-Gaussian energy and the relative entropy of non-Gaussianity are functionally equivalent, these states also maximize the relative entropy of non-Gaussianity at fixed energy, as previously identified in~\cite{non-gauss_2}. A simple example of such states are the Fock states $\ket{n}$, where $n\neq0$. 
This means that, under a fixed total energy budget, the class of pure states with zero first and second moment uniquely define the most non-Gaussian states.

In the same spirit, one may ask: given a class of non-Gaussian unitaries, which Gaussian input state yields an output state with maximum non-Gaussianity? As an example, consider the Selective Number-dependent Arbitrary Phase (SNAP) gate, which is energy-preserving~\cite{snap1,snap2,snap3}. 
SNAP gates, along with displacements, allow for arbitrary state generation in a bosonic cavity. For simplicity, we consider an input coherent state $\ket{\alpha_0}=\hat{D}(\alpha_0)\ket{0}$, and apply a SNAP gate that imparts a phase of $\pi$ to one of the Fock components $\ket{n}$.  
Without loss of generality, $\alpha_0$ is taken to be real and positive. The output state is given by
\begin{equation}
    \ket{\psi(\alpha_0,n)} = \hat{\Phi}_n\ket{\alpha_0},
\end{equation}
where $\hat{\Phi}_n = e^{i\pi \ket{n}\bra{n}}$.
The initial coherent state's energy only has a coherent contribution, \ie, $E\coh = |\alpha_0|^2$. From an energetic perspective, the SNAP gate redistributes the input coherent energy into output coherent, squeezed and non-Gaussian energy. For fixed $n$, the  optimization problem is to determine the value of $\alpha_0$ that maximizes the non-Gaussianity of $\ket{\psi(\alpha_0,n)}$. The problem reduces to finding the value of $\alpha_0$ that maximizes the non-Gaussian energy of the output state for each $n$.

\begin{figure}[ht]
    \centering
    \includegraphics[width=1\linewidth]{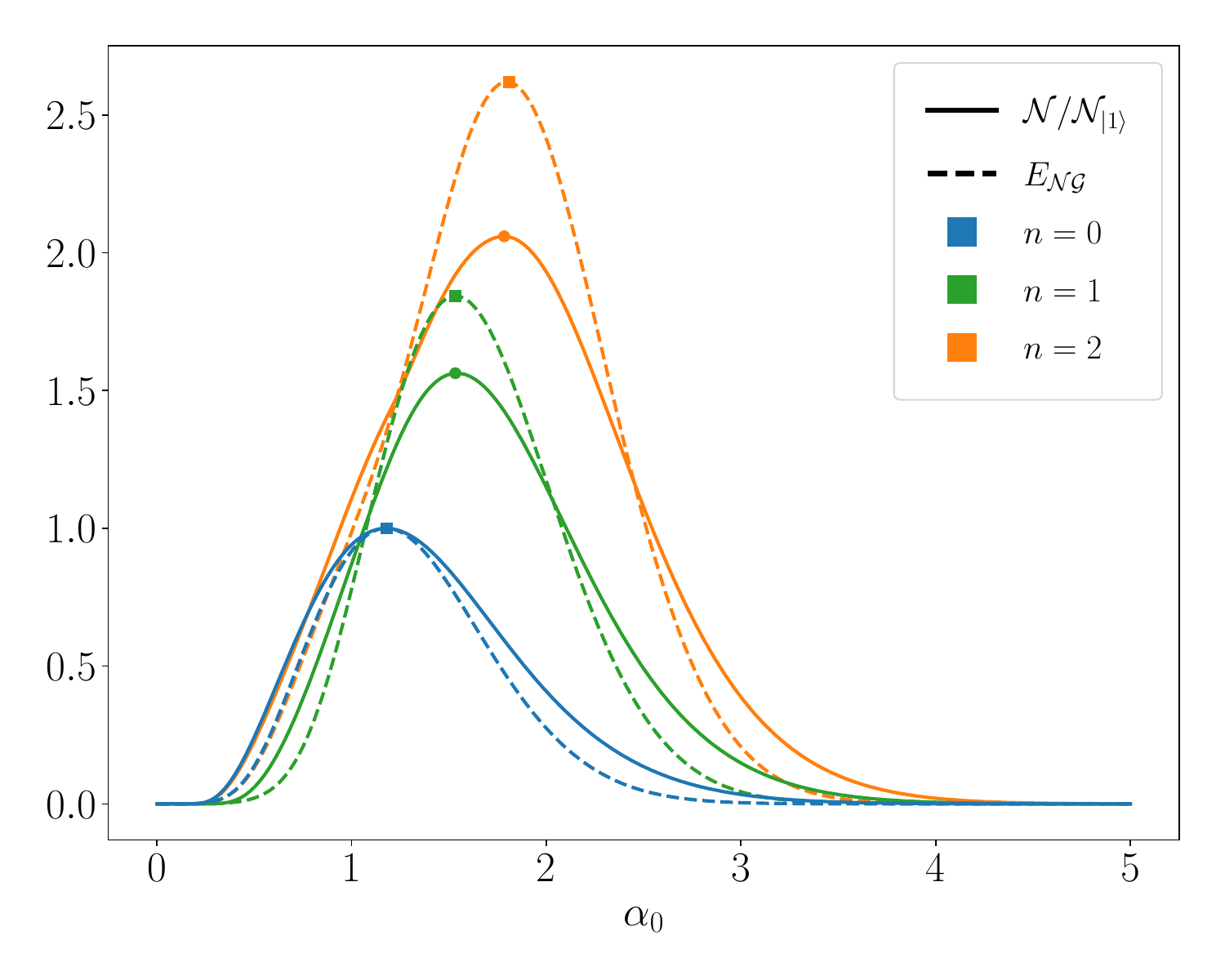}
    \caption{Wigner negativity (solid lines) and non-Gaussian energy (dashed lines)
        for the states $\ket{\psi(\alpha_0,n)} = \hat{\Phi}_n\ket{\alpha_0}$,
        with varying $\alpha_0$.
        We consider SNAP operations applied to the Fock components
        $n=0$ (blue), $n=1$ (green), and $n=2$ (orange). We observe that the optimal values of $\alpha_0$ that maximise the Wigner negativity closely match the values that maximise $E\ng$.}
    \label{pure_plot}
\end{figure}

In Fig.~\ref{pure_plot}, we plot the non-Gaussian energy as a function of $\alpha_0$ for different values of $n$. We observe that there exists an optimal value of $\alpha_0$
 that maximizes the non-Gaussian energy for each $n$. By tuning the input coherent amplitude to this value, one can generate the most non-Gaussian state within this setting. For comparison, we also plot the Wigner negativity of the output states as a function of $\alpha_0$ for the same values of $n$. Wigner negativity is another measure of non-Gaussianity for pure states and is given by~\cite{negativity_measure,non_gauss_0}
\begin{equation}
    \mathcal{N}[\rho] = \int d^2\alpha\, \big| W(\alpha) \big| - 1\,,
\end{equation} 
where
$    W(\alpha)
    :=
    \frac{2}{\pi}
    \Tr[
    \rho\,\hat{D}(\alpha)\hat{\Pi}\hat{D}^\dagger(\alpha)
    ]
$
is the Wigner function of the state and
$\hat{\Pi}=(-1)^{\hat{a}^\dagger\hat{a}}$ is the parity operator~\cite{wigner,Royer_parity_defn}.
We normalize Wigner negativities with respect to that of a single photon, \ie, ${\cal N}_{\ket{1}}:=\mathcal{N}[\op{1}{1}] = 4e^{-1/2}-2$. 

Firstly, we observe that the region where $E\ng > 0$ completely coincides with the region where ${\cal N} > 0$. This is due to Hudson's theorem~\cite{Hudson1974} which states that a pure non-Gaussian state is necessarily Wigner negative. Secondly, we observe that the values of $\alpha_0$
 that maximize the Wigner negativity closely coincide with those that maximize the non-Gaussian energy. This is noteworthy because distinct resource monotones generally need not induce the same ordering on states unless they are related by a monotonic function~\cite{entanglement_monotones_oredering}. While the Wigner negativity and the non-Gaussian energy are not explicitly related by a monotonic function, we nevertheless observe that their behavior is qualitatively similar, with their maxima occurring at remarkably similar values of $\alpha_0$. This observation suggests that, for pure states, non-Gaussian energy can act as an effective cost-function to maximise Wigner negativity, enabling the approximation of optimal parameters and reducing the search space. This is desirable because Wigner negativity is both analytically and computationally difficult to evaluate and optimise. In contrast, the non-Gaussian energy is significantly more tractable analytically and demands less computational resources, making it useful for optimization tasks.  Ref.~\cite{lim2026energetic} successfully applies this strategy to the scattering of a coherent pulse by a two-level atom. In addition to providing a more tractable metric for optimization, the non-Gaussian energy is more experimentally accessible than the negativity, the former only requiring knowledge of the first two moments which can be obtained through photon detection and homodyne detection and the latter requiring full state tomography of the output field. 




\section{Extension to mixed states}\label{sec:Extension to mixed states}
\subsection{Energetic decomposition for mixed states} 
We finally extend our energetic decomposition to mixed states of a single bosonic mode. As in the pure-state case, our goal is to identify an energetic quantity that can determine whether a given, possibly mixed, quantum state $\rho$ is Gaussian. A general mixed Gaussian state can be expressed as \begin{equation} \rho_G \equiv\rho_G(\alpha,\xi,\beta) = \hat{D}(\alpha)\hat{S}(\xi)\,\tau(\beta)\,\hat{S}\dg(\xi)\hat{D}\dg(\alpha)\,, \label{eq:gaussian_state_form_general} \end{equation} where $\tau(\beta) = e^{-\beta \ha\dg \ha}/\Tr[e^{-\beta \ha\dg \ha}]$ is a thermal state with temperature $\beta^{-1}$ (in units of $\hbar\omega$). \begin{figure}[H] \centering \includegraphics[width=1\linewidth]{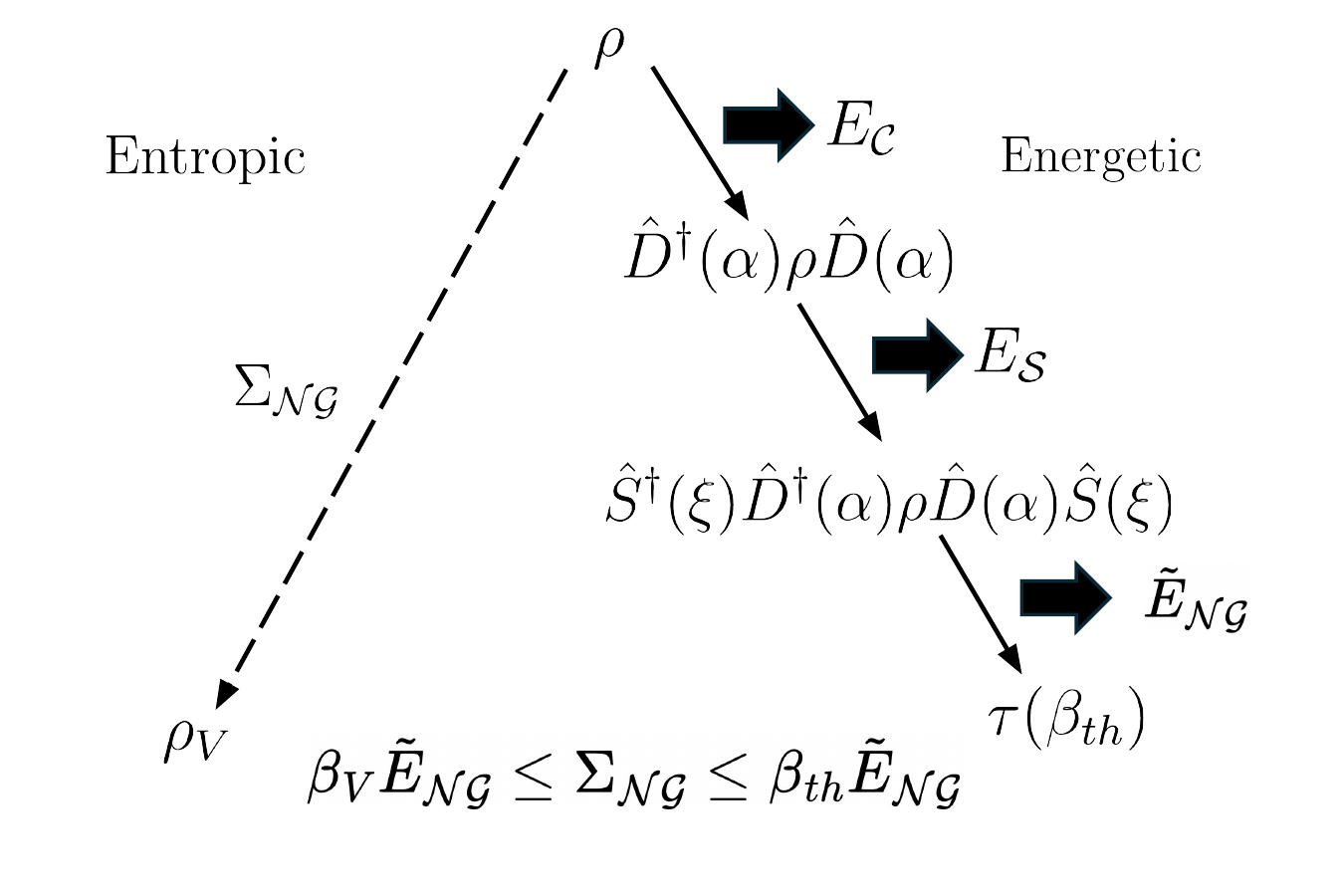} \caption{A diagrammatic representation of our energetic witness of non-Gaussianity compared with the entropic measure of non-Gaussianity. As in the pure-state case [Fig.~\ref{fig:enrgetic-diagram-pure}], our energetic approach first removes energy from the system via Gaussian unitaries. The remaining contribution is then extracted by transforming the state into the completely passive state. The entropic approach proceeds as in Fig.~\ref{fig:enrgetic-diagram-pure}.} \label{fig:energetic_diagram_mixed} \end{figure} For an arbitrary state $\rho$, we decompose the total energy as \begin{equation}\label{eq:mixed_decomp} E = E\coh + E\sq + E\mx + \ngmx, \end{equation} where $E\coh$ and $E\sq$ are given in Eqs.~\eqref{eq:coherent energy} and \eqref{eq:squeezing energy} and account for the displacement and squeezing contributions, respectively. To quantify the energetic contribution arising from the mixedness of the state, we introduce the \emph{thermal energy} $E\mx$, defined as \begin{equation} E\mx := n_{th}\,. \end{equation} Here, $n_{th}$ is the energy (in photon numbers) of a thermal state $\tau(\beta_{th})$ with temperature $\beta_{th}^{-1}$, chosen such that it has the same von Neumann entropy as $\rho$, namely $S[\tau(\beta_{th})]=S[\rho]$. The state $\tau(\beta_{th})$ is known as the completely passive state~\cite{complete_passive_Lenard1978, complete_passive_Pusz1978,Alicki}. For a Gaussian state $\rho_G$ of the form given in Eq.~\eqref{eq:gaussian_state_form_general}, $E\mx$ coincides with the energy of $\tau(\beta)$ and thus quantifies its thermal contribution. Therefore, the sum $E\coh+E\sq+E\mx$ accounts for the total energy of any Gaussian state. We then attribute the remaining contribution, \begin{equation} \label{eq:NG_energy_mixed_states} \ngmx := \sqrt{\det V} - \frac{1}{2}- n_{th} \,. \end{equation} to the non-Gaussianity of the state. We use the tilde notation to emphasize that $\ngmx$ is only one possible energetic quantity associated with non-Gaussianity, resulting from our choice to characterize the mixedness of the state through the entropy-equivalent thermal state. In phase-space terms, $\ngmx$ quantifies the extent to which $\sqrt{\det V}$, which characterizes the spread of the state, exceeds the contribution expected from a thermal state with the same entropy, namely $n_{th}+\tfrac{1}{2}$. It therefore captures the excess phase-space spread that cannot be attributed solely to statistical mixedness and instead reflects the non-Gaussian character of the state.

Just as in the pure state case, $E\coh$ can be extracted by a displacement unitary $\hat{D}(\alpha)$ that sets the first moment of $\rho$ to zero and $E\sq$ can be extracted by a squeezing unitary $\hat{S}(\xi)$ that sets the covariance matrix into its Williamson form. These operations will then bring $\rho$ to its corresponding Gaussian-passive state~\cite{marcus_passive,vittorio_work_extraction}, denoted by $\rho^p_  G$,
\begin{equation}
    \rho^p_G=\hat{S}\dg(\xi) \hat{D}\dg(\alpha)\rho  \hat{D}(\alpha) \hat{S}(\xi)\,. \label{eq:williamson theorem for general}
\end{equation}
It has been shown in~\cite{complete_passive_Lenard1978, complete_passive_Pusz1978, Alicki}
that if one further allows for arbitrary unitaries and global entangling unitaries on multiple copies of $\rho_G^p$, the energy of the state can at most be further reduced to that of the completely passive state, $\tau(\beta_{th})$.  $\ngmx$ corresponds to the amount of energy that can be further extracted from $\rho_G^p$ [see Fig.~\ref{fig:energetic_diagram_mixed}], \ie, 
\begin{equation}\label{eq:eng_gps}
\ngmx(\rho) = E[\rho_G^{p}] - E[\tau(\beta_{th})]\,.
\end{equation}
Note that for pure states, the completely passive state is the vacuum. In this case, $E\mx=0$ and $\ngmx$ reduces to the non-Gaussian energy as defined in Eq.~\eqref{eq:non-Gaussian energy_pure}, recovering the energetic decomposition for pure states [Eq.~\eqref{eq:energetic_decomposition_pure}].

\subsection{Faithful witness of non-Gaussianity}
We now show that $\ngmx$ provides a faithful energetic witness of non-Gaussianity. To this end, we exploit the positivity of the quantum relative entropy in Eq.~\eqref{relative enrtorpy-3}, which yields \begin{equation} S[\rho_V] \geq S[\rho]\,. \end{equation} Recall that $\rho_V$ denotes the reference Gaussian state. Equality holds if and only if $\rho$ itself is Gaussian. Using the expression for $S[\rho_V]$ given in Eq.~\eqref{eq:S of rhoV}, we obtain \begin{equation} h(\sqrt{\det V}) \geq S[\rho]\,, \end{equation} and therefore \begin{equation} \sqrt{\det V} \geq h^{-1}(S[\rho])\,. \label{eq:entropy_bound_mixed} \end{equation} For pure states, $S[\rho]=0$ and, since $h^{-1}(0)=\frac{1}{2}$, this reduces to the familiar Robertson-\sch{} uncertainty relation, $\sqrt{\det V} \geq \tfrac{1}{2}$. Equation~\eqref{eq:entropy_bound_mixed} may therefore be viewed as a mixed-state generalization of the Robertson-\sch{} bound, providing a tighter lower bound on $\det V$ that depends explicitly on the state's entropy. Gaussian states precisely saturate this bound. Using the relation $S[\rho] = S[\tau(\beta_{th})]=h(n_{th}+1/2)$, Eq.~\eqref{eq:entropy_bound_mixed} further implies that \begin{equation} \ngmx \geq 0\,. \end{equation} We can therefore summarize the key properties of $\ngmx$ as follows: 
\begin{enumerate} 
\item {\it Positivity}---It is a well-defined non-negative quantity, \ie, $ \ngmx\geq 0$. \label{prop:mix-1} 
\item {\it Faithfulness}---$\ngmx = 0$, iff $\rho$ is Gaussian.\label{prop:mix-2} 
\item {\it Invariance under Gaussian unitaries}---$\ngmx[\rho]=\ngmx[\hat{U}_G\rho \hat{U}_G^\dagger]$ for any Gaussian unitary $\hat{U}_G$. This follows from property~\eqref{prop:3_pure} of the non-Gaussian energy for pure states together with the fact that $n_{th}$, being solely determined by the von Neumann entropy, is invariant under unitary transformations.\label{prop:mix-3} \end{enumerate} 
Note that, for $\ngmx$ to qualify as a genuine measure of non-Gaussianity for mixed states, it must be non-increasing under Gaussian channels, \ie, $ \ngmx[{\cal G}(\rho)] \leq \ngmx[\rho], $ where $ {\cal G}(\rho)=\Tr_E [\hat U_G(\rho\otimes\sigma_E)\hat U_G^\dagger], $ $\sigma_E$ is a Gaussian environment state and $\hat U_G$ is a Gaussian unitary. For the important class of initially isotropic states (i.e., states with vanishing squeezing energy) evolving under thermalization channels, we prove in Appendix~\ref{appendix:Channel_Proof} that $\ngmx$ decreases monotonically. Nevertheless, this monotonicity property does not extend to arbitrary Gaussian channels. Indeed, Appendix~\ref{counterexample for eng mixed} provides an explicit counterexample showing that $\ngmx$ can increase under Gaussian dynamics.\\

\begin{figure}
    \centering
    \includegraphics[width=1\linewidth]{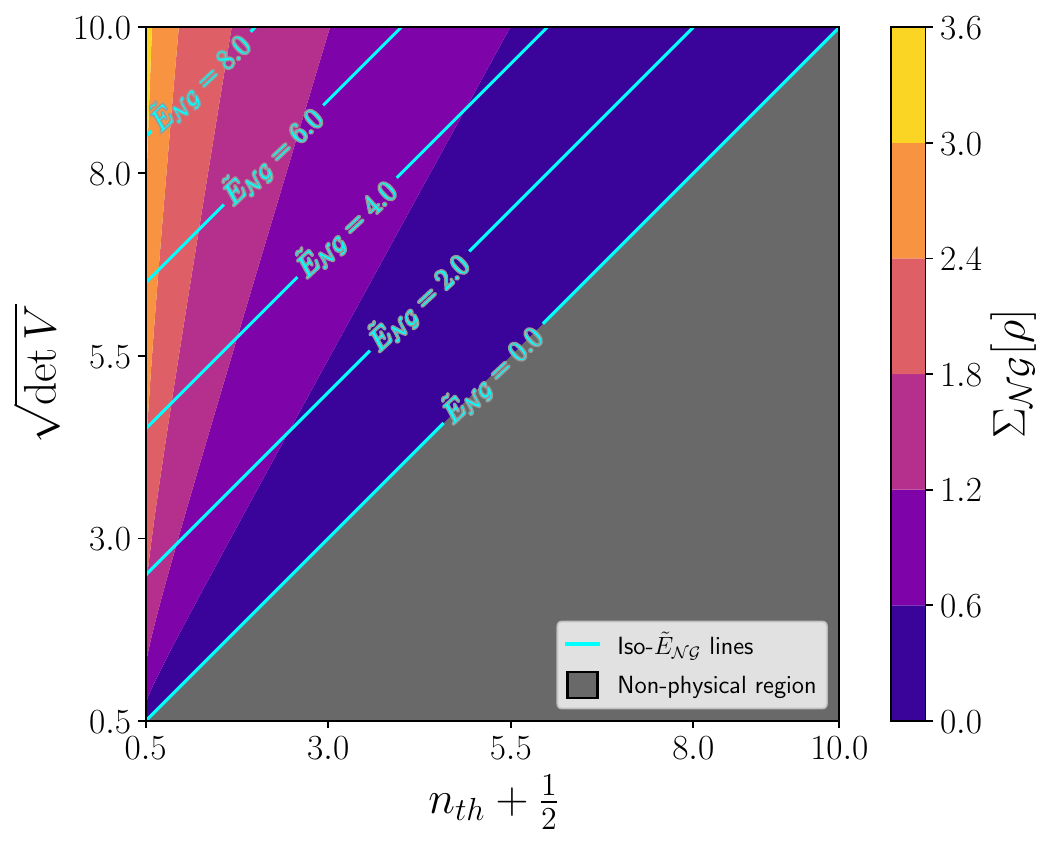}
    \caption{$\ngmx$ and  $\Sigma\ng$ as functions of $\sqrt{\det V}$ and $ n_{th}+\frac{1}{2}$. Physical quantum states only exist on and above the $\sqrt{\det V} = n_{th} + \half$ line. The cyan lines correspond to iso-$\ngmx$. The background color bar shows the variation of $\Sigma\ng$. All Gaussian quantum states lie on the $\sqrt{\det V} = n_{th} + \half$ line with the vacuum state at the origin of this plot, \ie, $(0.5,0.5)$. Pure non-Gaussian states lie on the vertical axis since $n_{{th}}=0$.   Qualitatively, both $\Sigma\ng$ and $\ngmx$  increase towards the upper left corner of the plot. However the constant lines of these two quantities are not parallel highlighting that these quantities are functionally independent.}
    \label{fig:iso reng and eng}
\end{figure}

\subsection{Visualisation and comparison with relative entropy of non-Gaussianity}
Properties~\ref{prop:mix-1}, \ref{prop:mix-2}, and \ref{prop:mix-3} establish $\ngmx$ as a faithful witness of non-Gaussianity, and we shall henceforth refer to it as the ``non-Gaussian witness''. It is instructive to visualize this quantity in the parameter space spanned by \[ (x_1,x_2)\equiv\left(\sqrt{\det V},\, n_{th}+\frac12\right), \] shown in Fig.~\ref{fig:iso reng and eng}. Property~\ref{prop:mix-1} implies that all physical states satisfy $x_1\geq x_2$ and therefore occupy the region above the line \[ \sqrt{\det V}=n_{th}+\frac12. \] A point in this diagram does not uniquely specify a quantum state; rather, all states sharing the same values of $\sqrt{\det V}$ and $n_{th}$ are represented by the same point. In particular, states related by Gaussian unitaries occupy the same location. The line $\sqrt{\det V}=n_{th}+\frac12$ corresponds precisely to the set of Gaussian states, for which $\ngmx=0$. The vacuum state is located at $(1/2,1/2)$, while pure non-Gaussian states lie on the vertical axis since $n_{th}=0$. Mixed non-Gaussian states occupy the interior of the physical region. Geometrically, $\ngmx=x_1-x_2$ measures the vertical distance from a given point to the Gaussian boundary, and its constant-value contours are shown by the cyan lines in Fig.~\ref{fig:iso reng and eng}. This representation also provides a natural framework for comparing the energetic witness $\ngmx$ with the relative entropy of non-Gaussianity. For an arbitrary mixed state $\rho$, the latter is given by [see Eq.~\eqref{relative enrtorpy-3}] \begin{equation}\label{eq:ngE_mixed_h_form} \Sigma\ng[\rho] = h(\sqrt{\det V}) - h\!\left(n_{th}+\frac{1}{2}\right)\,. \end{equation} Hence, both $\Sigma\ng[\rho]$ and $\ngmx$ depend only on the coordinates $(x_1,x_2)$ introduced above. Their gradients are \begin{equation} \nabla \Sigma\ng = \bigl(h'(x_1),-h'(x_2)\bigr)\,, \label{eq:grad-1} \end{equation} where $\nabla := \left(\frac{\partial}{\partial x_1},\frac{\partial}{\partial x_2}\right)$, while \begin{equation} \nabla \ngmx = (1,-1)\,. \end{equation} Since these vectors are generally not proportional, $\Sigma\ng[\rho]$ and $\ngmx$ are not monotonic functions of one another. Consequently, increasing $\ngmx$ does not necessarily imply an increase in $\Sigma\ng[\rho]$. This behavior is illustrated in Fig.~\ref{fig:iso reng and eng}, where contours of constant $\Sigma\ng[\rho]$ are shown together with contours of constant $\ngmx$. As expected, the two families of contours are not parallel. Nevertheless, both quantities exhibit a similar qualitative trend, increasing towards the upper-left region of the diagram corresponding to large $\sqrt{\det V}$ and small $n_{th}+\tfrac12$. Although $\ngmx$ and $\Sigma\ng[\rho]$ are not equivalent quantifiers for mixed states, they remain closely related. In Appendix~\ref{appendix:bound on increase of reng and eng}, we derive a necessary and sufficient condition for the relative entropy of non-Gaussianity to increase under an infinitesimal transformation $\rho\rightarrow\rho+\dd\rho$. Specifically, $\dd \Sigma\ng[\rho]>0$ if and only if \begin{equation} \dd\ngmx > \left(\frac{\beta_{th}}{\beta_V} - 1 \right) \dd E\mx\,. \label{increase_of_reng} \end{equation} Here, $\dd\ngmx = \ngmx[\rho+\dd\rho] - \ngmx[\rho]$, $\dd E\mx = E\mx[\rho+\dd\rho] - E\mx[\rho]$, and $\beta_V$ denotes the inverse temperature of the Gaussian-passive state associated with $\rho_V$. Furthermore, in Appendix~\ref{appendix: Proofof bounds}, we prove that $\ngmx$ provides both upper and lower bounds on $\Sigma\ng[\rho]$: \begin{equation} \beta_V \ngmx \leq \Sigma\ng[\rho] \leq \beta_{th}\ngmx\,. \label{eq:energetic_bounmds} \end{equation} These inequalities establish a quantitative connection between the energetic and entropic characterizations of non-Gaussianity for mixed states. The bounds coincide only for Gaussian states, for which $\Sigma\ng[\rho]=\ngmx=0$.

\section{Conclusion}
In this work, we introduced energetic quantities that characterize the non-Gaussianity of single-mode bosonic states. For pure states, we established a decomposition of the total energy into coherent, squeezing, and non-Gaussian contributions. We showed that the non-Gaussian contribution, which we termed the non-Gaussian energy, defines a bona fide measure of non-Gaussianity and elucidated its connection to the relative entropy of non-Gaussianity. Moreover, unlike many existing measures of non-Gaussianity, the non-Gaussian energy can be accessed experimentally without requiring full state tomography. For mixed states, we identified an energetic contribution associated with non-Gaussianity by removing the thermal energy that accounts for the state's mixedness. We showed that this quantity provides a faithful witness of non-Gaussianity. Furthermore, we established necessary and sufficient conditions for the increase of the relative entropy of non-Gaussianity in terms of our energetic quantities and derived upper and lower bounds on the relative entropy of non-Gaussianity from the non-Gaussian witness. More broadly, our results reveal a close connection between energetic concepts and information-theoretic quantifiers of non-Gaussianity. By providing an energetic framework for its characterization, our work offers new tools for the detection, quantification, and efficient generation of non-Gaussian states in single-mode bosonic systems. 

An important direction for future research is the construction of an energetic measure of non-Gaussianity for mixed states that is monotonic under Gaussian operations, as well as its extension to general multimode bosonic systems. Another promising avenue is to investigate the interplay between quantum resources such as entanglement, squeezing, and non-Gaussianity through a unified energetic perspective. Finally, a deeper understanding of the role played by non-Gaussianity in the performance of continuous-variable quantum engines and in the cooling of quantum systems~\cite{cooling_valerio} may benefit from the energetic framework developed here.

\acknowledgments{The authors warmly thank Zacharie Van Herstraeten and Hugh Coleman for enlightening discussions and valuable comments. In particular, we are grateful to them for providing counterexamples showing that the non-Gaussian witness is not monotonic under Gaussian operations. This project is supported by the National Research Foundation, Singapore through the National Quantum Office, hosted in A*STAR, under its Centre for Quantum Technologies Funding Initiative (S24Q2d0009), by the National Research Foundation through the CNRS\@ CREATE internal grant ``NGAP'' (NRF2023-ITC004-001), and by the Plan France 2030 through the projects NISQ2LSQ (Grant ANR-22-PETQ-0006), OQuLus (Grant ANR-22-PETQ-0013), and OECQ through BPI France. This research is supported by the National Research Foundation, Prime Minister's Office, Singapore under its Campus for Research Excellence and Technological Enterprise (CREATE) programme.


%

\appendix

\section{The non-equivalence of Gaussian-passive state and core state}\label{appendix:core state}
The Gaussian-passive state is not necessarily a  core state~\cite{stellar_rank}. A state $\ket{\psi}=\sum_{n\geq 0}c_n\ket{n}$
is a core state if its stellar function is a polynomial, that is, 
\begin{equation}
    \ket{\psi}=F_{\psi}^{*}(a^\dagger)\ket{0},
\end{equation}
where $F_{\psi}^{*}(\alpha)
    =\sum_{n\geq 0}c_n\frac{\alpha^n}{\sqrt{n!}},
    \quad \forall\,\alpha\in\mathbb{C}$
is the stellar function of the state~\cite{stellar_rank}.
For example, consider the state
\[
\ket{\phi}=\frac{1}{\sqrt{2}}(\ket{0}+\ket{1}).
\]
Its stellar function is
\[
F_{\phi}^{*}(\alpha)
=\frac{1}{\sqrt{2}}(1+\alpha),
\]
which is a polynomial. Hence, $\ket{\phi}$ is a core state. However, $\ket{\phi}$ is not Gaussian-passive, since its coherent energy is nonzero,
$E\coh=\frac{1}{4}$.
The Gaussian-passive state of $\ket{\phi}$, obtained by removing the coherent energy is 
\begin{equation}
    \hat{D}\left(-\frac{1}{2}\right)\ket{\phi}
    =
    F^*_{\hat{D}(-1/2)\ket{\phi}}(a^\dagger)\ket{0},
\end{equation}
where $F^*_{\hat{D}(-\frac{1}{2})\ket{\phi}}(\alpha)= \frac{e^{-1/8}}{\sqrt{2}}
e^{-\alpha/2} 
\left(\alpha+\frac{3}{2}\right)$ is the stellar function and is not a polynomial. Thus, the Gaussian-passive state is not necessarily a core state.  

\section{Proof that the non-Gaussian witness is non-increasing under thermalization channel for isotropic states}\label{appendix:Channel_Proof}
Here we show that $\ngmx$ is non-increasing under thermalization channels $\mathcal{G}_{th}$ for the set of state with zero squeezed energy. Consider a Gaussian state $\rho$ with zero coherent and squeezed energy undergoing thermalization with a bath having $\bar{n}$ thermal excitations. Let the spontaneous emission rate of $\rho$ be $\gamma$ and the energy of its reference Gaussian state $\rho_{V}$ be $E_{V}=\sqrt{\det V}-1/2 = \langle \delta a^{\dagger} \delta a\rangle$. Under the action of the thermalization channel, $E_{V}$ evolves as 
\beq\label{eq:E_V_evo}
E_{V}(t) = e^{-\gamma t}E_{V} + (1-e^{-\gamma t})\bar{n}\,. 
\eeq
We now use the fact that among all input states with a fixed input entropy, the output entropy of a thermalization channel is minimized by a thermal input state \cite{DeGio2017PRL}. This implies the inequality 
\beq\label{eq:ent_Gth_ineq}
S[\mathcal{G}_{th}(\tau(\beta_{th}))] \leq S[\mathcal{G}_{th}(\rho_V)]\,,
\eeq
where $\tau(\beta_{th})$ is the completely passive state of $\rho$ that describes its thermal energy $E\mx$. After interacting for time $t$, $\mathcal{G}_{th}$ transforms the thermal excitations of $\tau(\beta_{th})$ to $e^{-\gamma t}E\mx + (1-e^{-\gamma t})\bar{n}$, whereas the mixed energy of $\rho$ becomes $E\mx(t)$. As entropy is a monotonic function of the thermal excitations, from the entropy inequality \erf{eq:ent_Gth_ineq}, it follows that 
\begin{equation}
    e^{-\gamma t}E\mx + (1-e^{-\gamma t})\bar{n} \leq E\mx(t).
\end{equation}
Subtracting $E_{V}(t)$ from the above inequality and using \erf{eq:E_V_evo}, we find
\begin{equation}
    E_{V}(t) - E\mx(t) \leq e^{-\gamma t} (E_{V} - E\mx).
\end{equation}
As $0\leq e^{-\gamma t} \leq 1 $ and $\ngmx[\rho] = E_{V} - E\mx$, we arrive to the result $\Delta \ngmx = \ngmx[\mathcal{G}_{th}(\rho)] - \ngmx[\rho] = (E_{V}(t) - E\mx(t)) - (E_{V} - E\mx) \leq 0$, \ie, $W\ng$ decreases under the action of the Gaussian channel ${\cal G}_{th}$. When the state $\rho$ also has coherent energy, then the proof follows similarly as the coherent energy also decays exponentially as $e^{-\gamma t}$.

\section{Example of a Gaussian channel where the non-Gaussian witness increases}\label{counterexample  for eng mixed}
We provide an example where the non-Gaussian witness, $\ngmx$, increases under a Gaussian channel. Consider the quantum-limited amplifier channel~\cite{weedbrook}, $\mathcal{G}_{la}(\rho)$, which has the following Stinespring dilation:
\begin{equation}
   \mathcal{G}_{la}(\rho) =
   \Tr_E\!\left[
   \hat{U}_G \left(\rho\otimes \op{0}{0}_E\right) \hat{U}_G^\dagger
   \right],
\end{equation}
where
\begin{equation}
   \hat U_G =
   e^{r(\hat{a}^\dagger \hat{b}^\dagger-\hat{a}\hat{b})}
\end{equation}
is the two-mode squeezing unitary with the squeezing parameter $r>0$. Here, $\hat{a}$ and $\hat{b}$ denote the annihilation operators of the system and environment modes, respectively.

 Consider the initial Fock state $\rho=\op{1}{1}$. Its initial non-Gaussian contribution is
\begin{equation}
   \ngmx(\op{1}{1})=1.
\end{equation}
For a squeezing parameter $r=1$, the output state has
\begin{equation}
   \ngmx\!\left(\mathcal{G}_{la}(\op{1}{1})\right)\approx1.13.
\end{equation}
Thus, the above example demonstrates that $\ngmx$ can increase under a Gaussian channel.

\section{Proof of inequality~\eqref{increase_of_reng}}
\label{appendix:bound on increase of reng and eng}
Let us consider an infinitesimal operation on the state $\rho$ which changes the state from $\rho\rightarrow\rho+\dd\rho$. Beginning from \erf{eq:ngE_mixed_h_form}, and simplifying the notation so that let $x_1=\sqrt{\det V}$ and $x_2=n_{th}+\frac{1}{2}$, we can take the total differential the relative entropy if non-Gaussianity, yielding $\dd \Sigma\ng[\rho] = h'(x_1)\dd x_1 - h'(x_2)\dd x_2$. For the relative entropy of non-Gaussianity to increase, \ie, $\dd \Sigma\ng[\rho] > 0$, it must be the case that $h'(x_1)\dd x_1 > h'(x_2)\dd x_2$. Rearranging this inequality gives
\begin{equation} \label{eq:simp_ineq}
(\dd x_1-\dd x_2)> \left(\frac{h'(x_2)}{h'(x_1)}-1\right)\dd x_2\,.
\end{equation}
The left-hand side of this equation is simply the change in non-Gaussian energy is  $\dd \ngmx  =(\dd x_1-\dd x_2)$, while the change in thermal energy is $\dd E\mx=\dd x_2$. 
Furthermore, using the relation $h'(x)=\beta$, where $\beta$ denotes the inverse temperature of a thermal state whose symplectic eigenvalue is $x$, we identify $h'(\sqrt{\det V})=\beta_V$ and $h'(2n_{th}+1)=\beta_{th}$. Substituting these expressions into \erf{eq:simp_ineq} yields
\begin{equation}\label{condition_for_increase_of_entropy}
\dd \ngmx\ge \left(\frac{\beta_{th}}{\beta_V}-1\right)\dd E\mx\,.
\end{equation}

\section{Proof of Eq.~\eqref{eq:energetic_bounmds}} \label{appendix: Proofof bounds}
Let us begin by first showing that $\Sigma\ng[\rho] \leq \beta_{th} \ngmx$. Focussing on the left-hand side, we have, by definition,
\beq
\beta_{th}\inv \Sigma\ng[\rho] = \beta_{th}\inv (S[\rho_V] - S[\rho])\,.
\eeq
Since the state $\tau(\beta_{th})$ has the same entropy as $\rho$, \ie, $S[\rho] = S[\tau(\beta_{th})]$ and using the fact that the entropy of a thermal state is 
\beq\label{eq:ent_th}
S[\tau(\beta)] = \beta E[\tau(\beta)] + \ln Z(\beta)\,, 
\eeq
where $Z(\beta) = \Tr[e^{-\beta\ha\dg\ha}]$, we have
\beq
\beta_{th}\inv \Sigma\ng[\rho] = \beta_{th}\inv (S[\rho_V] - \ln Z(\beta_{th})) - E\mx\,,
\eeq
where we have recognised that $E[\tau(\beta_{th})] = E\mx$. By definition, we have that $S[\rho_V] = S[\tau(\beta_V)]$ and using the fact that the relative entropy 
\beq\label{eq:rel_ent_thermal}
D[\rho||\tau(\beta)] = \beta E[\rho] - S[\rho] + \ln Z(\beta)\,,
\eeq
we have
\beq\label{eq:pre_ineq1}
\beta_{th}\inv \Sigma\ng[\rho] = \ngmx - \beta_{th}\inv D[\tau(\beta_V)||\tau(\beta_{th})]\,.
\eeq
The right-hand side follow by noticing that $E[\tau(\beta_V)] = E[\rho_G^p]$, which can be seen from the fact that $E[\rho_V] = E[\rho]$ and the unitaries that transform $\rho_V \to \tau(\beta_V)$ also transform $\rho\to \rho_G^p$. Thus, $E[\tau(\beta_V)] - E\mx = \ngmx$. Finally, since $D[\rho_V||\tau(\beta_{th})] \geq 0$, we have
\begin{equation}
    \Sigma\ng[\rho] \leq \beta_{th} \ngmx\,.
\end{equation}
For the other bound, $\beta_V\ngmx \leq \Sigma\ng[\rho]$, let us consider \erf{eq:eng_gps}. Using the fact that $E[\rho_G^p] = E[\tau(\beta_V)]$ as well as \erf{eq:ent_th}, we have
\beq
\ngmx = \beta_V\inv S[\tau (\beta_V)] - \ln Z(\beta_V) - E[\tau(\beta_{th})]\,.
\eeq
Making use of \erf{eq:rel_ent_thermal}, and that $S[\tau(\beta_V)] = S[\rho_V]$ and $S[\tau(\beta_{th})] = S[\rho]$, we obtain
\beq
\ngmx = \beta_V\inv (S[\rho_V] - S[\rho]) - \beta_V\inv D[\tau(\beta_{th})||\tau(\beta_{V})]\,.
\eeq
Finally, once again using the fact that $D[\tau(\beta_{th})||\tau(\beta_{V})]\geq 0$, we have
\begin{equation}
    \beta_V\ngmx \leq \Sigma\ng[\rho]
\end{equation}
 The lower and upper bounds of Eq.~\eqref{eq:energetic_bounmds} coincide only when the state is Gaussian, \ie, when $\Sigma\ng = \ngmx = 0$. Note that for a Gaussian state, $\tau(\beta_{{th}})=\tau(\beta_V)$.

\end{document}